# High efficiency dielectric metasurfaces at visible wavelengths

Robert C. Devlin[1], Mohammadreza Khorasaninejad[1], Wei-Ting Chen[1], Jaewon Oh[1,2] and Federico Capasso[1*]

[1]John A. Paulson School of Engineering and Applied Sciences, Harvard University, 9 Oxford Street, McKay 125, Cambridge, Massachusetts 02138, USA

[2]University of Waterloo, Waterloo, ON N2L 3G1, Canada

**Metasurfaces are planar optical elements that hold promise for overcoming the limitations of refractive and conventional diffractive optics[1-3]. Dielectric metasurfaces demonstrated thus far[4-10] are limited to transparency windows at infrared wavelengths because of significant optical absorption and loss at visible wavelengths. It is critical that new materials and fabrication techniques be developed for dielectric metasurfaces at visible wavelengths to enable applications such as three-dimensional displays, wearable optics and planar optical systems[11]. Here, we demonstrate high performance titanium dioxide dielectric metasurfaces in the form of holograms for red, green and blue wavelengths with record absolute efficiency (>78%). We use atomic layer deposition of amorphous titanium dioxide that exhibits low surface roughness of 0.738 nm and ideal optical properties. To fabricate the metasurfaces we use a lift-off-like process that allows us to produce highly anisotropic nanofins with shape birefringence. This process is applicable to any optical element and can be implemented using a broad class of materials.**

The ability to realize high efficiency dielectric metasurfaces (DMs) is critically dependent on the optical properties of the constituent material, as characterized by the complex refractive index, $\tilde{n} = n + ik$. The material should have negligible absorption loss ($k \approx 0$) with relatively high refractive indices ($n > 2$) at visible wavelengths. While a negligible absorption is necessary for high transmission efficiency, a high refractive index ensures strong confinement of the light that ultimately allows full control over the phase of the exiting wavefront (phase changes from 0 to 2 $\pi$) over nanoscale dimensions. The material should be optically smooth, having a surface roughness that is much smaller than the wavelength of light in the material. High surface roughness introduces extraneous light scattering, which is a significant source of loss. To minimize roughness, the material must be amorphous or single crystalline, as the grain boundaries in polycrystalline materials increase roughness and act as light scattering centers.

We chose amorphous $TiO_2$ as our base material because $TiO_2$ has a transparency window ($k \approx 0$) for wavelengths longer than ~350 nm and its interband transition lies just outside of the visible spectrum, resulting in a sufficiently high index of refraction for strong light-matter interactions. We deposit $TiO_2$ by atomic layer deposition (ALD) at 90° C with tetrakis(dimethylamido)titanium (TDMAT) as the precursor due to its high deposition rate and absence of defect-driven absorption that is typical of $TiCl_4$- based precursors[12]. Additionally, use of the ALD process offers several advantages[13]. ALD is a self-limiting process providing conformal coverage and monolayer control of the film thickness. The conformal nature of the technique is essential to produce high aspect-ratio $TiO_2$ nanostructures. The uniform and monolayer coverage yields smooth films with minimal defects compared to those prepared via sputtering or evaporation. Furthermore, the ALD process also allows us to precisely control the phase of deposited $TiO_2$, producing amorphous, rutile, or anatase films depending on the deposition temperature[14].

Characterization of the optical properties of our ALD TiO2 show that it is an ideal material for DMs. Figure 1a shows the measured optical properties of a $TiO_2$ film from ultraviolet (243 nm) to near infrared (900 nm) wavelengths. To extract the $TiO_2$ optical properties from the collected data, we used a Tauc-Lorentz (TL) oscillator model developed for amorphous materials[15]. Over the visible spectrum the refractive index ranges from 2.63 to 2.34 and remains relatively flat between $\lambda$ = 500-750 nm ($\Delta n$ = 0.09). Below the wavelength of 500 nm the index of refraction increases rapidly, implying the approach of an electronic transition. For wavelengths shorter than 360 nm, the imaginary part of the refractive index, $k$, begins to take on non-zero values, a result of interband absorption. The inset of Fig. 1a shows a plot of the square root of the absorption coefficient $\left(\sqrt{\alpha} = \sqrt{4\pi k/\lambda}\right)$. We use this plot to extract the bandgap of our $TiO_2$ through linear extrapolation to $\alpha$ = 0, giving $E_g$ = 3.45 eV. Similarly, $E_g$ is a fitting parameter in the TL model and is 3.456 eV (the full set of fitting parameters are included in the supplementary information [SI]). This value of the bandgap is in good agreement with previously reported values for amorphous $TiO_2$[16].

Our $TiO_2$ films also exhibit surface roughness that is much smaller than the incident wavelength, as characterized by atomic force microscopy (AFM). Figure 1b shows an AFM scan of a typical $TiO_2$ film deposited on a fused silica substrate. The film has a root mean square (RMS) roughness of 0.738 nm, which is limited by the surface roughness of the underlying substrate (Supplementary Fig. 1). Due to the absence of grain boundaries in AFM, coupled with the X-ray diffraction data (Supplementary Fig. 2), we conclude that the material is amorphous as expected. As is the case with the optical bandgap, this value of roughness for our amorphous $TiO_2$ is of the same order of magnitude as previously reported values, which is an order of magnitude less than other phases such as polycrystalline rutile or anatase[14,17,18]. The latter two phases generally have grain boundaries and RMS roughness as high as 5 – 10 nm both of which contribute to light scattering loss[19].

To achieve highly efficient metasurface devices while preserving the optical properties of ALD-prepared $TiO_2$, we use the fabrication process[20] shown in Fig. 2. We spin electron beam resist (EBR) onto a fused silica substrate to produce a layer with thickness, $t_{resist}$ (Fig. 2a). (The spin speed of the resist controls the thickness). Control of $t_{resist}$ is important because it sets the height of our final nanostructures. We pattern the resist using electron beam lithography and subsequent development in solution to remove the exposed EBR. This pattern is the inverse of our final metasurface (Fig. 2b). We transfer the exposed sample to an ALD chamber set to 90° C. The purpose of this temperature is twofold: it produces the desired amorphous phase and keeps the EBR below its glass transition temperature (deterioration of the nanoscale patterns). During the deposition, the gaseous $TiO_2$ precursor (TDMAT) coats all exposed surfaces, producing a conformal film on the top and side of the EBR as well as on the exposed fused silica substrate (Fig. 2c). We allow the ALD process to reach a specific thickness (or equivalently proceed for a sufficient number of cycles) such that all features have been completely filled with $TiO_2$. Since the conformal ALD process fills the gaps from both sides the total ALD film thickness required is $t_{film} \geq w/2$, where w is the maximum width of all gaps (Fig. 2d). In practice we allow the deposition to proceed well beyond the minimum requirement of half the feature width to ensure that $TiO_2$ has sufficiently diffused into all pores and that there are no air voids in the final nanostructures. We remove the residual $TiO_2$ film that coats the top surface of the resist by reactive ion etching of the sample in a mixture of $BCl_3$ and $Cl_2$ gas (8:2), similar to a planarization technique. The etch depth is equal to $t_{film}$ so that the etching process exposes the underlying resist and the top of the nanostructures (Fig. 2e). We remove the remaining resist and leave only the nanostructures that make up our metasurface (Fig. 2f). In this way, we obtain structures of heights $t_{resist}$ while needing only to deposit a film of thickness $t_{film} \approx w/2$, which is timesaving and efficient.

We note that this approach is different from standard liftoff techniques, as one cannot employ conventional liftoff process due to the conformal coating of the ALD films. In addition, creating anisotropic nanostructures via dry etching of $TiO_2$, similar to other dielectric material, is difficult and can lead to chemical contamination and increased sidewall roughness. The use of ALD for infiltrating nanoscale structures has been previously used[17,21] but the patterns created are generally fixed by a set template (e.g., anodic alumina or inverse opal) whereas here we are free to define more complex nanostructures because we deposit directly onto exposed electron beam resist.

Investigation of the structures, fabricated using the process above, shows that we produce nanostructures (nanofins) with the desired anisotropy and nanoscale dimensions. Figure 3 shows SEM images of a fabricated metasurface hologram. As seen in Fig. 3a, we can densely pattern large areas with subwavelength spaced $TiO_2$ nanofins. This ultimately ensures that the majority of the incident light is imprinted with a desired phase, that higher diffraction orders are suppressed and that we can produce, in our case, holographic images with high efficiency. Figure 3b shows an SEM of the metasurface at high magnification, allowing resolution of individual nanofins. With this process we can also obtain extremely small spacing between individual nanofins, as shown by the 6 nm gap in Fig. 3b. The structures in this case have dimensions of 250 nm x 90 nm but we can produce structures with dimensions as small as 40 nm (Fig 3c). It is clear that the residual EBR has been removed after the process and only clean, independent nanofins remain.

Structures that deviate from 90$^o$ sidewalls, taking on a more triangular shape in cross section, will cause the effective index to change as light propagates through the metasurface and introduce errors in the designed phase. Similarly, structures with large voids or defects will change the index of the nanofins. Figure 3d shows a cross section SEM of the nanofins (vertical direction in the image is the direction of light propagation). We observe that the slope of the structures in the vertical direction is 89$^o$, that is, the nanostructures are highly anisotropic. This anisotropy is despite the fact that the nanostructures shown here are relatively tall compared to their other dimensions, with heights of 600 nm. Similarly, we do not see any void formation in the center of the nanofins.

To demonstrate the efficiency and functionality of our $TiO_2$ metasurfaces we designed three metasurface holograms (meta-holograms) to have peak efficiencies at wavelengths of 480 nm, 532 nm and 660 nm. The building blocks of meta-holograms are the birefringent nanofins shown in Fig. 3b and c, in which we impart the required phase via nanofin rotations—known as geometric or Pancharatnam-Berry (PB) phase[22-24]. (For more detail on this technique see refs.[4,24,25], methods and SI). An advantage of the PB phase is that the imparted phase is wavelength independent (only the efficiency changes with wavelength), thus providing a unique platform to test the performance of our metasurface over the entire visible range. We compute the phase map of a holographic image, the binary Harvard logo, by means of the Gerchberg-Saxton phase-retrieval algorithm[26] (Supplementary Fig. 3).

Figure 4a to 4c show the measured and simulated efficiencies as a function of wavelength from 480 nm to 800 nm for the meta-holograms designed at $\lambda$ = 480 nm, 532 nm and 660 nm, respectively. We define absolute efficiency is as the total intensity of the reconstructed Harvard logo divided by the transmitted intensity through a 300 x 300 $\mu m^2$ square aperture. (See methods and Supplementary Fig. 4 for measurement details). The experimental results generally follow the simulation data and reach maxima of 82%, 81% and 78% near the design wavelengths of 480 nm, 532 nm and 660 nm, respectively. These values are, to our knowledge, the highest reported to date even compared to reflective metasurfaces[27-29]. We note that in certain cases there are discrepancies between the simulated and measured trend. Differences between designed and fabricated nanofin dimensions and the possibility of weak coupling between fins likely cause these discrepancies. Moreover, the meta-hologram designed at 660 nm has broadband efficiency (>60% from 550 nm to 710 nm), and the reconstructed images remain high quality throughout the visible range. Figure 4d to 4i show the holographic images across the visible spectrum that the TiO$_2$ DMs generate (see Supplemental Video 1 for full set of images). The subwavelength spacing and oversampling of the phase map can be seen in the images since there is sharp resolution of fine features such as the word "VERITAS" at the top of the Harvard crest. The bright spot near the edge of the cross of the Harvard logo is from the zero-order. However at the design wavelength the ratio of the intensity in the zero order to that of hologram image is 1%.

While we have chosen to demonstrate our process using Panchartnam-Berry phase meta-holograms, the TiO$_2$ properties and fabrication process are not limited to this specific type of metasurface. For example, simulations using the measured optical constants of our TiO$_2$ and structural dimensions achievable with our fabrication process show that we can vary pillars dimensions in order to provide full 2 $\pi$ phase coverage without using geometric phase accumulation (Supplementary Fig. 6). Thus one can also use the results we presented here to produce DMs that use linear birefringent resonators to encode phase information[8]. Moreover, our process and resulting metasurfaces enable the implementation of any high efficiency DMs components at visible wavelengths such as axicons, lenses and polarizers.

We have detailed the first experimental realization of highly efficient dielectric metasurfaces that spans the visible spectrum. We employed ALD to produce smooth amorphous $TiO_2$ films that are transparent for wavelengths longer than 360 nm and have an index of refraction that is sufficiently high to provide complete phase control over an optical wavefront. The fabrication technique for metasurfaces, requiring only a single step of lithography, provides a simple method to produce the highly anisotropic nanostroctures that are necessary for DMs and while we used $TiO_2$, our process is applicable to any material that can be deposited via ALD. The fabricated metasurface holograms have the highest recorded efficiencies to-date (82%, 81% and 78%) at their respective design wavelengths. These metasurface also exhibited broadband efficiency due to their implementation via the geometric phase. However, the technique presented here is general and can be applied to any type of metasurface. Careful consideration of the optical properties of our base material and the precision of the fabrication technique allowed us to expand the functionality of DMs to visible wavelengths. This work can enable creation of new compact optical systems at visible wavelengths with thicknesses that are orders of magnitude less than traditional optical systems.

**Acknowledgements**

This work was supported by the Air Force Office of Scientific Research (AFOSR) FA9550-14-1-0389 MURI grant. R.C.D is supported by a fellowship through Draper Laboratory. This work was performed in part at the Center for Nanoscale Systems, a member of the National Nanotechnology Infrastructure Network, which is supported by the National Science Foundation under NSF award no. ECS-0335765. CNS is part of Harvard University.

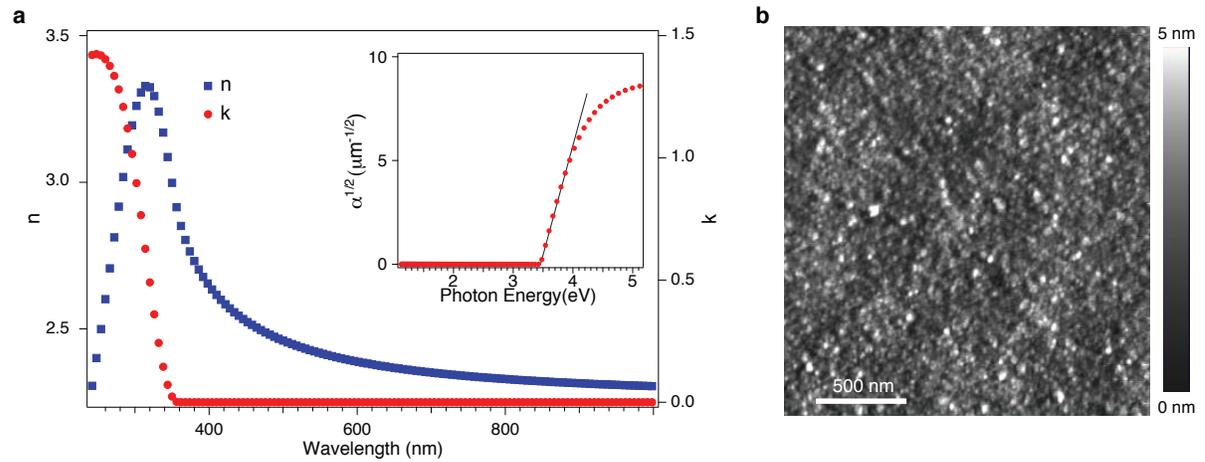

**Figure 1 | Amorphous titanium dioxide material properties**. **a**, Measured real (blue squares) and imaginary (red circles) part of the refractive index (*n* and *k*) as a function of wavelength across the visible spectrum. Inset shows the square root of the absorption coefficient as a function of energy, which we use to obtain the bandgap of 3.456 eV. **b**, Atomic force microscope image of a typical $TiO_2$ film deposited via atomic layer deposition. The film is atomically smooth surface with RMS roughness of 0.738 nm

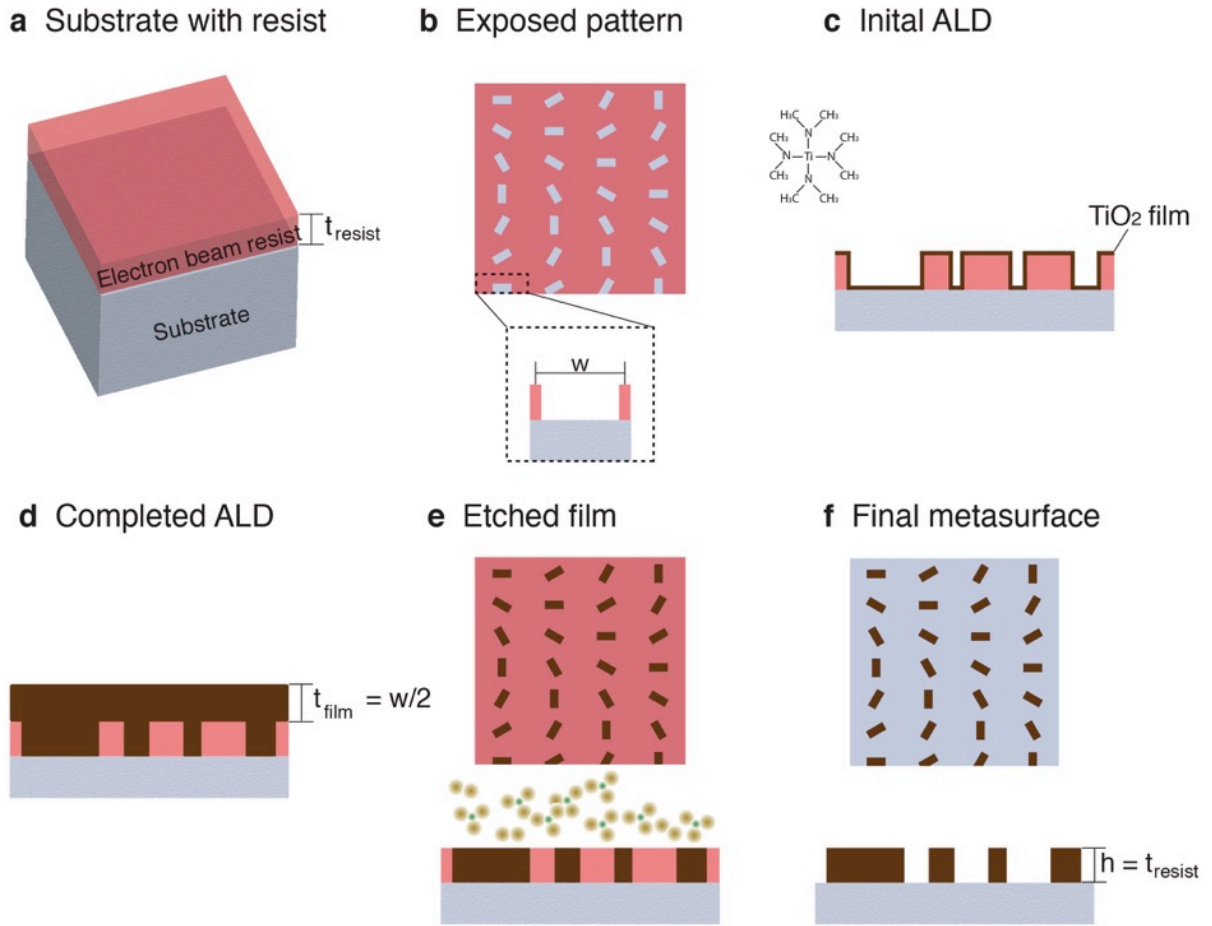

**Figure 2 | Fabrication process for dielectric metasurfaces**. **a**, Electron beam resist (EBR) on fused silica (or equivalent transparent substrate) with thickness $t_{resist}$ that ultimately sets the height of the final structure (perspective view). **b**, Inverse of the final metasurface pattern imprinted into the EBR by electron beam lithography and subsequent development of the pattern (top view). The boxed area is an expanded cross section of the maximum feature width, w. **c**, Initial $TiO_2$ deposition via atomic layer deposition conformally coats sidewalls and top of the EBR and exposed substrate (side view). TDMAT molecule used for atomic layer deposition is also shown. **d**, Completed deposition of the $TiO_2$ yields a thickness of the film greater than half the width of the maximum feature size, $t_{film} \geq w/2$. **e**, Exposed tops of $TiO_2$ metasurface and residual EBR after reactive ion etching with a mixture of $Cl_2$ and $BCl_3$ ions (top and side view). **f**, Final dielectric metasurface after removal of remaining EBR (top and side view)

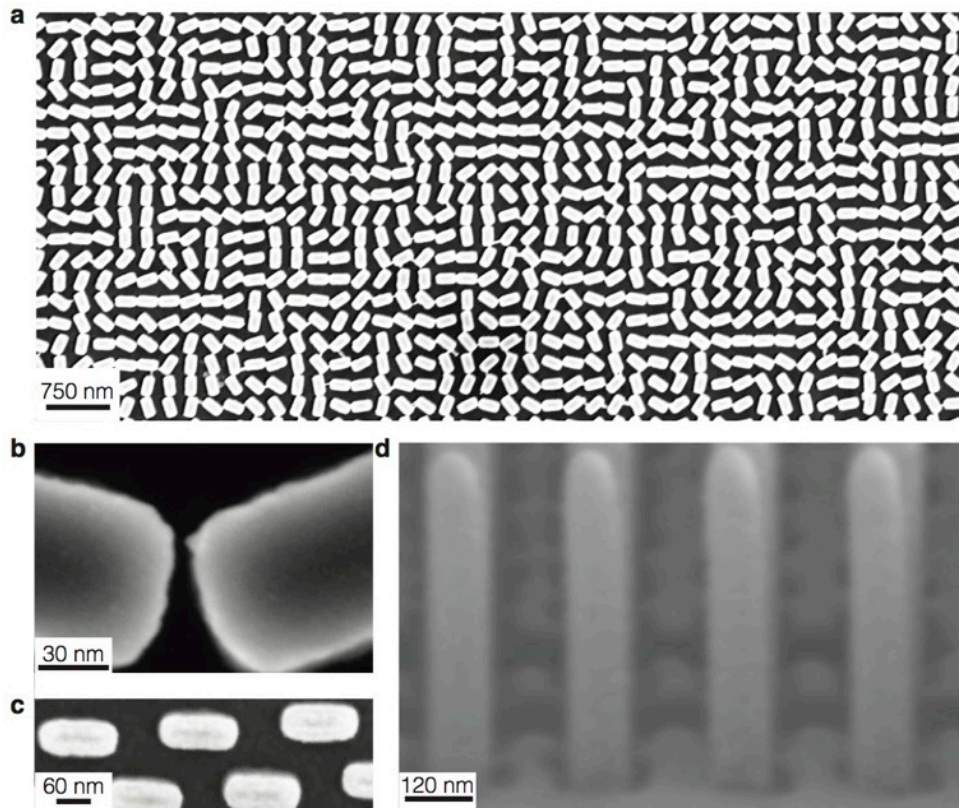

**Figure 3| Scanning electron microscope images of fabricated structures**. **a**, Large scale view of a fabricated metasurface hologram composed of $TiO_2$ nanofins. **b**, Zoomed top-view of metasurface showing individual nanofins. The fabricated nanofins are free from residual resist and have dimensions ±10 nm of the designed dimensions of 250 nm x 85 nm. It can also be seen that with this fabrication technique we can achieve gaps between structures as small as 6 nm. **c**, Top view of structures with lateral dimensions of approximately 40 nm **d**, Cross section (side-view) of nanofins exhibiting vertical sidewalls with a height of approximately 600 nm. The contrast oscillations between nanofins results from shadowing effects during deposition of a metal film that prevents charging while we image the samples.

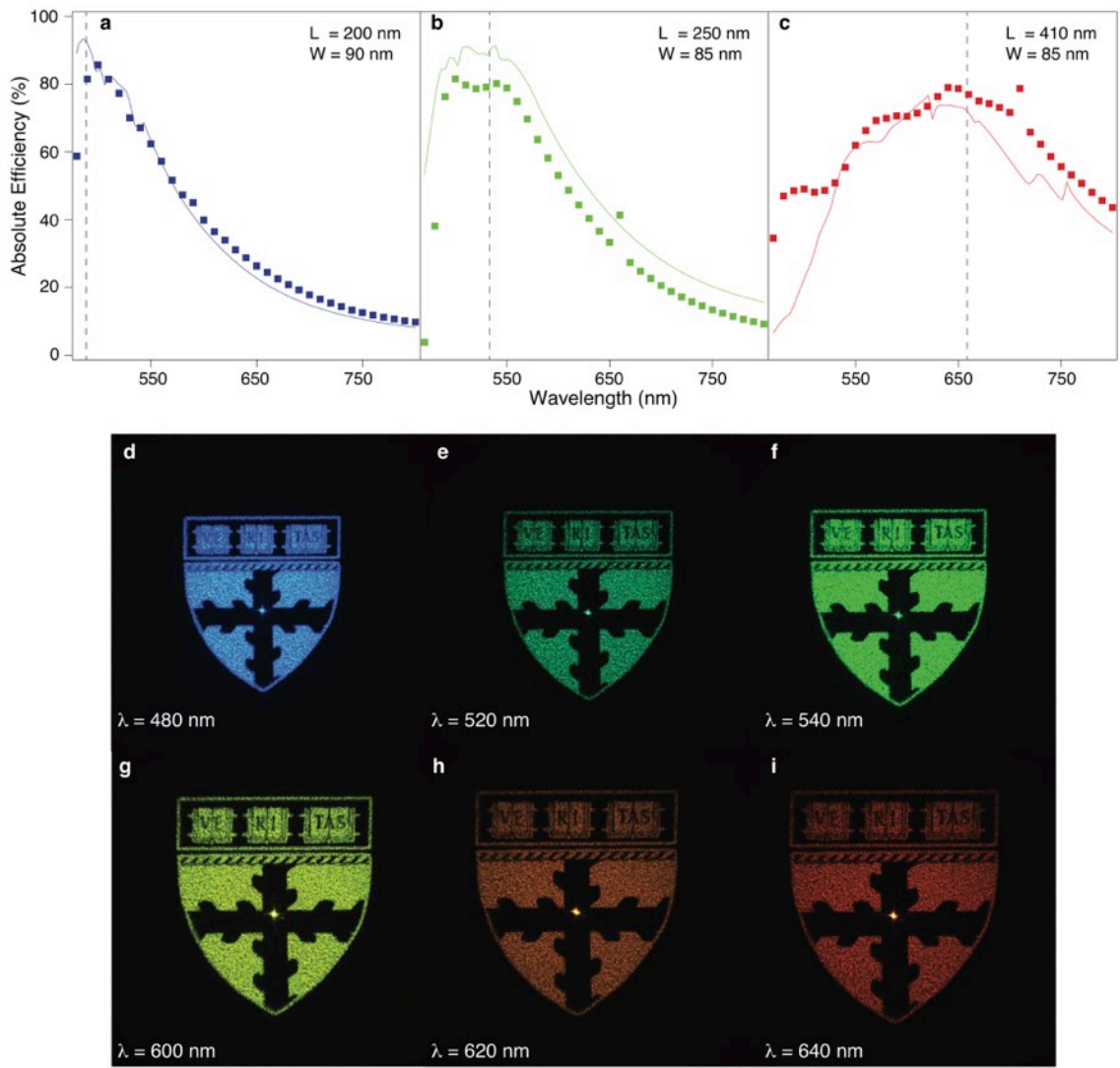

**Figure 4 | Measured absolute efficiencies and holographic images. a-c**, Measured (square markers) and simulated (solid line) hologram efficiencies. We define the absolute efficiency as the ratio of the total intensity of the hologram to the total intensity transmitted through an aperture of the same size as the hologram (300 x 300 mm$^2$). The vertical dashed line marks the design wavelengths of the each device and device dimensions are given in the top right of each panel. **d-i**, Holographic images covering the visible spectrum. The input wavelength is shown in the bottom of each panel. All images were obtained from the device designed for 480 nm and show the broadband behavior of a single device. The bright spot in the center of the image is due to the propagation of zero-order light.